\documentclass[twocolumn]{rmf-d}

\usepackage{epsf}
\usepackage{fancyhdr}
\usepackage{rmfbib}

\newif\ifpreprint
\preprinttrue

\newcommand{\rmfcornisa{}}

\begin{document}

\twocolumn[
\title{
\ifpreprint
\protect\rightline{\large UASLP--IF--02--001}
\vskip0.3cm
\fi
Simulation of a RICH Detector for the CKM Experiment
}

\author{
{Ibrahim Torres$^1$, J\"urgen Engelfried$^2$, Antonio Morelos$^3$}
}
\address{
Instituto de F\'{\i}sica, Universidad Aut\'onoma de San Luis Potos\'{\i},
\'Alvaro Obregon 64, Zona Centro, San Luis Potos\'{\i}, SLP 78000, M\'exico
\\
email: $^1$itorres@ifisica.uaslp.mx, $^2$jurgen@ifisica.uaslp.mx, 
$^3$morelos@ifisica.uaslp.mx
}
\date{Janury 30, 2002} 
\maketitle

\begin{abstract}
We will present here the simulations of RICH detectors which will
be used in the CKM experiment. We will verify their performance,
critical to the experiment.
\keys{
RICH detector; phototubes; chromatic dispersion;  rare kaon decay
}
\end{abstract}

\begin{resumen}
En este trabajo, se presenta una simulaci\'on con el fin de verificar el
desempe\~no de dos detectores RICH que se usar\'an en el experimento CKM.\\
\descript{
Detector RICH; fototubos; decaimiento raro de kaones; dispersi\'on crom\'atica
}
\end{resumen}

]

\ifpreprint
\thispagestyle{fancy}
\pagestyle{fancy}
\fancyhead{ }
\fancyfoot[L]{Submitted to {\sl Revista Mexicana de F\'{\i}sica}}
\fancyfoot[C]{\thepage}
\fancyfoot[R]{ }
\renewcommand{\headrulewidth}{0pt}
\fi

\section{Introduction}
In this article we present results on simulating a RICH
detector which will be used within the CKM experiment\cite{CKM},
a rare kaon decays experiment recently approved at Fermilab.
The main objective is to verify if the resolution of the current design
of the detector 
is sufficient for the CKM experiment.

After a short description of the CKM experiment, we will detail the
requirements for the detector system, describe the current design,
our method of simulation, and the results obtained so far\cite{tesis}.

\section{Description of the CKM Experiment}
CKM stands for ``Charged Kaons at the Main Injector''.
The main goal of this experiment is to measure the branching ratio
of $K^{+} \rightarrow \pi^{+} \nu \bar{\nu}$ to about $10\,\mbox{\%}$
statistical precision. This measurement can be used to extract 
in a theoretically clean way the
magnitude of $V_{td}$, an element of the
Cabibbo Kobayashi Maskawa matrix,
with an overall precision of $10\,\mbox{\%}$\cite{1prop}.

This measurement plays an important role in testing the description
of ${\cal CP}$ violation in the Standard Model, and is complementary
to the current efforts of the $B$-factories

The prediction of the Standard model\cite{1prop} for the branching
ratio of $K^+\rightarrow \pi^+\nu\bar\nu$ is
$B(K^+ \to \pi^+ \nu \bar\nu)=0.75 \pm 0.29\times 10^{-10}$
and was first calculated by Inami and Lim\cite{7prop}.

The E787 experiment at Brookhaven, with a stopped Kaon beam, recently
observed two clean events\cite{brookhaven}, leading to a measured
branching ratio of
$B=1.57^{+1.75}_{-0.82}\cdot 10^{-10}$. 
A follow-up experiment at Brookhaven, E949,
will exploit fully the stopped kaon beam technique, and expects to observe
5-10~events.
 
To reach the 100~event level, CKM will use kaon decays in flight.
Due to kinematic backgrounds, the acceptance will only be in the
$1-2\,\mbox{\%}$ range, and this, together with the spill structure of the
Main Injector ($1\,\mbox{s}$ out of $3\,\mbox{s}$) 
and a running time of 2 years,
requires at least 30,000,000~kaons passing the detector per second.
This places high performance criteria on the detectors.
With an expected branching ratio of $10^{-10}$ and 100~signal events with
less than 10~background events to observe, the experiment has the challenge
of suppressing all sources of possible backgrounds to the $10^{-12}$ level.

The most obvious backgrounds are the main $K^+$ decays modes:
$K^{+} \rightarrow \mu^{+} \nu_\mu\,\, (B=64\,\mbox{\%})$ and 
$K^{+} \rightarrow \pi^{+} \pi^0\,\, (B=21\,\mbox{\%})$.
In the first, miss-identifying the $\mu^+$ as $\pi^+$, and in
the second not registering the two photons from the $\pi^0$ decay
lead to identical topologies like the signal $K^+\to\pi^+\nu\bar\nu$.

We will describe here the experimental method on the $\pi^+\pi^0$ mode,
other modes work the same.
With a very efficient Photo-Veto-System\cite{secondproposal} 
CKM expects to reject this
mode by $10^{-7}$. For the additional rejection down to the $10^{-12}$ level
we use kinematic rejection: By measuring the invariant mass of the
system the $\pi^+$ recoils ({\sl the missing mass}), we obtain a
peak at the $\pi^0$ mass in the $\pi^+\pi^0$ two-body decay, and a continuous
distribution in the signal mode (a three-body decay). By excluding the 
mass region around the $\pi^0$, we obtain our signal events.
The missing mass $M_{miss}$ is given (in a very good approximation) by
\begin{displaymath}
M^{2}_{miss} = m^2_K (1- p_\pi/p_K) + m_\pi^2 (1- p_K/p_\pi) - p_\pi p_K \theta^2
\end{displaymath}
where $m_K$ is the mass of the $K^+$, $m_\pi$ is the mass of the $\pi^+$,
$p_K$ and $p_\pi$ are their corresponding momenta, and $\theta$ is the
decay angle. These are the only experimentally accessible quantities
in the signal decay.

CKM will use two different types of spectrometer systems to measure
the momentum vectors of the $K^+$ and the $\pi^+$, a magnetic
spectrometer with high-rate MPWCs  and straw-tubes,
and a velocity spectrometer\cite{13exp},
a novel concept,
consisting of two Ring Imaging Cherenkov counters\cite{14exp}. 
The resolutions
in momentum and angle of these detectors has to be good enough to
allow for a narrow enough missing mass peak of the $\pi^0$ not to
overlap too much with the signal region.  It also requires a
magnetic field homogeneity of better than a few per mil.

\section{Requirements and Design Criteria for the Velocity Spectrometer}

The velocity spectrometer has to measure independently of the
conventional magnetic spectrometer the vector velocity of
the incoming kaon and the outgoing pion.  In a first approximation
for designing the experiment\cite{first_proposal}
we assumed momentum resolutions of $0.5\%$ and $1\%$ respectively.

The RICH for measuring the outgoing $\pi^+$ was modeled after
the very successful phototube based SELEX RICH detector\cite{selex}, 
keeping the diameter of the vessel about the same, but doubling
the vessel length to $20\,\mbox{m}$ and correspondingly
the radius of curvature of the spherical mirror to $40\,\mbox{m}$.
As radiator gas we will use Neon, as in the SELEX RICH.
The phototubes are sensitive down to $160\,\mbox{nm}$, requiring
also sufficient reflectivity of the mirrors in the VUV.

The Kaon RICH, to measure the velocity vector of the incoming $K^+$,
and also to discriminate against other beam particles (mostly $\pi^+$),
had to be designed differently. We will still keep the main feature
from the SELEX RICH, the use of photomultipliers for the detection
of the Cherenkov photons.

\begin{figure}[htb]
\begin{center}
\leavevmode
\epsfxsize=\hsize
\epsfbox{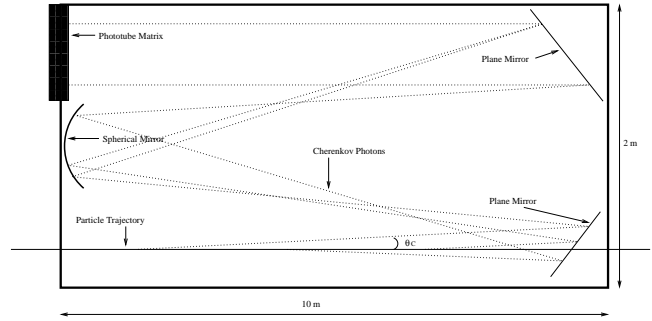}
\caption{Principle design of the Kaon RICH for CKM used in this study}
\label{krch}
\end{center}
\end{figure}

The beam in CKM is very parallel,
and has a diameter of only $10\,\mbox{cm}$.  Since the detector
is located just before
the decay volume, we have to avoid interactions in detector material,
asking for as less as possible amount of material in the beam.  To
reduce the effect of the phototube size on the overall resolution,
we also decided to double the focal length, leading to a doubled
ring radius. All this together lead us to a design shown in
fig.~\ref{krch}.
The vessel has a length of $10\,\mbox{m}$, but we
double-fold the light-path, with a (thin) plane mirror in the
beam, a spherical mirror of $40\,\mbox{m}$ radius ($20\,\mbox{m}$ focal
length) and another plane mirror, both outside the beam.

\begin{figure}[htb]
\begin{center}
\leavevmode
\epsfxsize=\hsize
\epsfbox{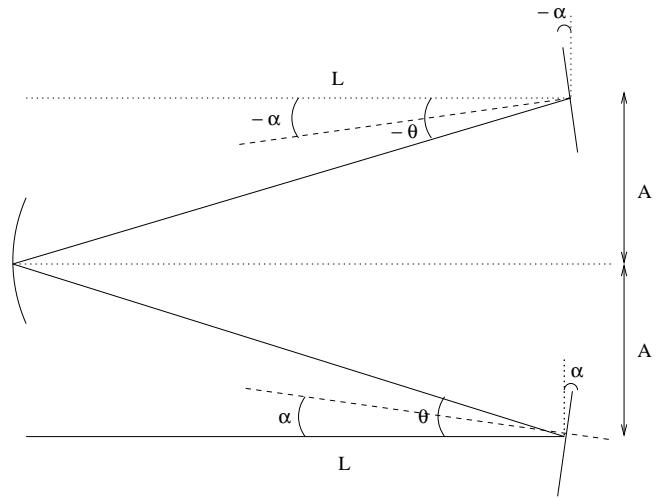}
\caption{Detailed layout of the mirror geometry for the Kaon RICH.} 
\label{espejos}
\end{center}
\end{figure}

A detailed design of the mirror arrangement is shown in fig.~\ref{espejos}.
The two plane mirrors have the same distance from the spherical mirror,
as well horizontally as vertically, so that the tilt-angles of the
two plan mirrors have the same absolute value.  This assures normal incidence
on the photo-tube plane without tilting it. 
With simple geometry it can be seen that the inclination angle $\alpha$
for the plane mirrors is given by
\begin{displaymath}
\alpha={{1}\over{2}}\theta = {{1}\over{2}} \arctan\left( {{A}\over{L}}\right)
\end{displaymath}
where the focal length of the spherical mirror is given by $F=L+L/\cos\theta$.

The size of the mirrors can be determined from the
Cherenkov angle, given by\cite{cherenkov}
\begin{equation}
\cos\theta_C= \frac{1}{\beta n}\label{eqn:cherenkov}
\end{equation}
where $\beta = v/c$ is the velocity of the particle as a fraction
of the velocity of light in vacuum, and $n$ is the refractive index
of the medium.
The sizes of the 3 mirrors are determined assuming a $K^+$ with
a momentum of $22\,\mbox{GeV}/c$, and the size of the 
photocathode (a matrix of $1/2\,\mbox{inch}$ photomultipliers)
is determined assuming a $\pi^+$ with the same momentum.
The sizes obtained depend on the refractive index $n$ of the
radiator gas. In table~\ref{tab:tamanos} we show typical values
\begin{table}[htb]
\begin{center}
\begin{tabular}{|l|l|}\hline
 & Diameter\\ \hline
Thin plane mirror & $43.5\,\mbox{cm}$\\
Second plane mirror& $90.5\,\mbox{cm}$\\
Spherical mirror & $79.5\,\mbox{cm}$\\
Phototube array & $90.5\,\mbox{cm}$\\ \hline
\end{tabular}
\caption{Sizes for the mirrors and the phototube array for
${\rm CF}_4$ at $0.8\,\mbox{atm}$.
\label{tab:tamanos}
}
\end{center}
\end{table}
for the final parameters obtained in this study.

\section{Data Analysis and Results}

After coding the full geometrical description of all relevant materials
within the GEANT\cite{geant} package, and incorporating all the routines into
the standard experiment Monte Carlo program, we studied as a first application
the momentum resolution of the detector as a function of pressure for
two different radiator gases, namely ${\rm N}_2$ and ${\rm CF}_4$.
${\rm N}_2$ was selected because it's refractive index at $1\,\mbox{atm}$
of $n-1\approx 300\cdot10^{-6}$ gives the correct threshold for kaons
of $22\,\mbox{GeV}/c$, but we were afraid of chromatic dispersion
(change of $n$ as a function of wavelength\cite{nitrogen}).
For this reason we
selected also ${\rm CF}_4$, with lower dispersion\cite{CF4}, 
at the cost of running at a non-atmospheric pressure.

Using a $22\,\mbox{GeV}$ $K^+$, where GEANT takes care of all possible
interactions like multiple scattering etc.,\ including the production
and tracking of Cherenkov photons  and simulation of the wavelength 
depending efficiency function of the phototubes\cite{hamamatsu}, 
we obtained a list
of hit phototubes for every $K^+$ we simulated.  For every event, we
performed a simple ring-fit\cite{ringfit} to determine
the radius of the Cherenkov ring, obtaining at the end an average
radius~$r$ and its standard deviation~$\sigma_r$.
The radius~$r$ in a RICH as function of the momentum~$p$ for a particle
with mass~$m$ is given by (with a small-angle approximation of 
eq.~\ref{eqn:cherenkov})
\begin{equation}
r(p)=\frac{R}{\sqrt{2}}\left(
1-\frac{1}{n(1-m^2/p^2)^\frac{1}{2}}\right)^\frac{1}{2}
\end{equation}
and its derivative by
\begin{equation}
\frac{dr}{dp}=\frac{R}{2\sqrt{2}}\left(
  1-\frac{(1-m^2/p^2)^{-\frac{1}{2}}}{n}\right)^{-\frac{1}{2}}\left(\frac{1}{n}\left( 1-\frac{m^2}{p^2}\right)^{-\frac{3}{2}}(m^2/p^3)\right)  
\label{dp}
\end{equation}
where $R$ denotes the radius of curvature of the spherical mirror.
With the simple relation
$\sigma_p=\frac{dp}{dr}\sigma_r$
we can determine the momentum resolution $\sigma_p$ for any set of parameters
we wish to simulate.

\begin{figure}[htb]
\begin{center}
\leavevmode
\epsfxsize=\hsize
\epsfbox{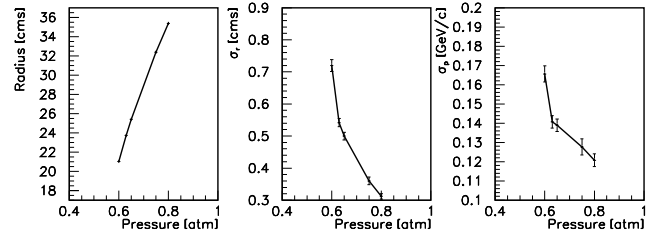}
\caption{Example results for values obtained from the simulation with
${\rm CF}_4$ as radiator gas, for
$K^+ $ with $22\,\mbox{GeV}/c$, all as a function of pressure.
Left: Cherenkov Ring Radius. Middle: Ring Radius Resolution $\sigma_r$.
Right: Momentum Resolution $\sigma_p$.}
\label{sigma_p}
\end{center}
\end{figure}

As an example we show in fig.~\ref{sigma_p} 
the radius~$r$, the radius resolution~$\sigma_r$,  and the momentum 
resolution $\sigma_p$ as a function of the pressure for ${\rm CF}_4$ as
radiator gas. An increase in pressure corresponds to an increase in 
the refractive index $n$ of the radiator gas.

\begin{figure}[htb]
\begin{center}
\leavevmode
\epsfxsize=\hsize
\epsfbox{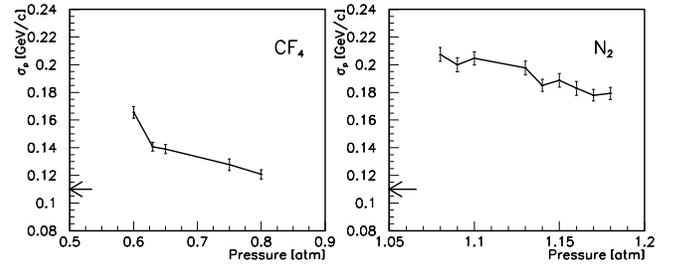}
\caption{Results obtainted with the  GEANT simulation for
the momentum resolution $\sigma_p$ for the Kaon RICH, with
(left) ${\rm CF}_4$ and (right) ${\rm N}_2$ as radiator gas.
The arrows indicate the proposed momentum resolution of
$0.5\,\mbox{\%}$ for a $22\,\mbox{GeV}/c$ $K^+$.}
\label{resfinal}
\end{center}
\end{figure}

In fig.~\ref{resfinal} we show the final result of this study, the momentum
resolution~$\sigma_p$ as a function of pressure for two possible
radiator gases, ${\rm CF}_4$ and ${\rm N}_2$. The arrow indicates
the required momentum resolution of $0.5\,\mbox{\%}$ for a $22\,\mbox{GeV}/c$
$K^+$.  As seen, the required resolution can be obtained by using
${\rm CF}_4$ as radiator gas.  As shown in table~\ref{contrib}, we 
analysed the different contributions to the resolution. The main
factor is the chromatic dispersion.
\begin{table}[htb]
\begin{center}
\begin{tabular}{|l|c|}
\hline
\multicolumn{1}{|c|}{Type}&$\sigma_r$\\
\hline
Multiple Scattering & $0.09\,\mbox{cm}$\\
Chromatic Dispersion & $0.49\,\mbox{cm}$\\
\hline
Total & $0.49\,\mbox{cm}$\\
\hline
\end{tabular}
\caption{Contributions to the ring radius resolution of the 
Kaon RICH with ${\rm CF}_4$ at $0.65\,\mbox{atm}$ as radiator gas.
\label{contrib}
}
\end{center}
\end{table}

\section{Conclusions}
In this work we performed a GEANT simulation of a RICH detector to
determine its response and momentum resolution as a function
of pressure for two different radiator gases. 

We obtained as a result that with ${\rm CF}_4$ at $0.8\,\mbox{atm}$ pressure
as radiator gas we can achieve the required resolution of $0.5\%$.

The programs and routines written for this study were used extensively
for other design studies for the second edition of the proposal
for the CKM experiment\cite{secondproposal}.  The experiment
was approved by the Fermilab Director in June~2001.  We are now
improving and testing the design of the RICH detectors in preparation
for a Technical Design Report.

\section{Acknowledgment}
We would like to thank Peter Cooper and Erik Ramberg
for fruitful discussions,
and the other members of the CKM Collaboration for continuous 
and encouraging support.
This work was supported by CONACyT-Mexico under Grant~28435-E and by
FAI-UASLP.

\ifpreprint
\def\refname{\bf References}
\fi



\end{document}